\documentclass[12pt,preprint]{aastex}
\usepackage{graphicx}

\shorttitle{Comments on MG11}
\shortauthors{L\'opez-Corredoira et al.}

\begin{document}

\title{Comments on the paper `Unifying Boxy Bulge and Planar Long Bar in the Milky Way' by Mart\'\i nez-Valpuesta \& Gerhard}

\author{M. L\'opez-Corredoira\altaffilmark{1,2}, A. Cabrera-Lavers\altaffilmark{1,3}, 
C. Gonz\'alez-Fern\'andez\altaffilmark{4}, F. Garz\'on\altaffilmark{1,2}, T. J. Mahoney\altaffilmark{1}, J. E. Beckman\altaffilmark{1,2,5}}
\affil{$^1$ Instituto de Astrof\'\i sica de Canarias, E-38200 La Laguna, 
Tenerife, SPAIN\\
$^2$ Departamento de Astrof\'\i sica, Universidad de la Laguna, E-38206 La Laguna, 
Tenerife, SPAIN\\
$^3$ GTC Project Office, E-38205 La Laguna, Tenerife, SPAIN\\
$^4$ Departamento de F\'\i sica, Ingenier\'\i a de Sistemas y Teor\'\i a de la Se\~nal,
Universidad de Alicante, Apdo. 99, E-03080, Alicante, SPAIN\\
$^5$ Consejo Superior de Investigaciones Cient\'\i ficas, SPAIN}


\altaffiltext{1}{{\it e-mail}: martinlc@iac.es}

\begin{abstract}
We comment on the recent paper by Mart\'\i nez-Valpuesta \& Gerhard (2011), who suggest, as an alternative to the bulge + long bar hypothesis in the inner 4 kpc of our Galaxy, a single boxy-bulge structure with a twisted
major axis. In principle, we find this proposal acceptable; indeed, from a purely morphological point of view, this is more a question of semantics than
science, and possibly all of us are talking about the same thing. 
However, we think that the particular features of this new proposal of a ``single twisted bulge/bar'' 
scenario leaves certain observational facts unexplained, whereas the model of a misaligned bulge + long bar
successfully explains them.
\end{abstract}

\section{Discussion}

\begin{figure}
\vspace{1cm}
\hspace{1cm}\resizebox*{8cm}{8cm}{\includegraphics{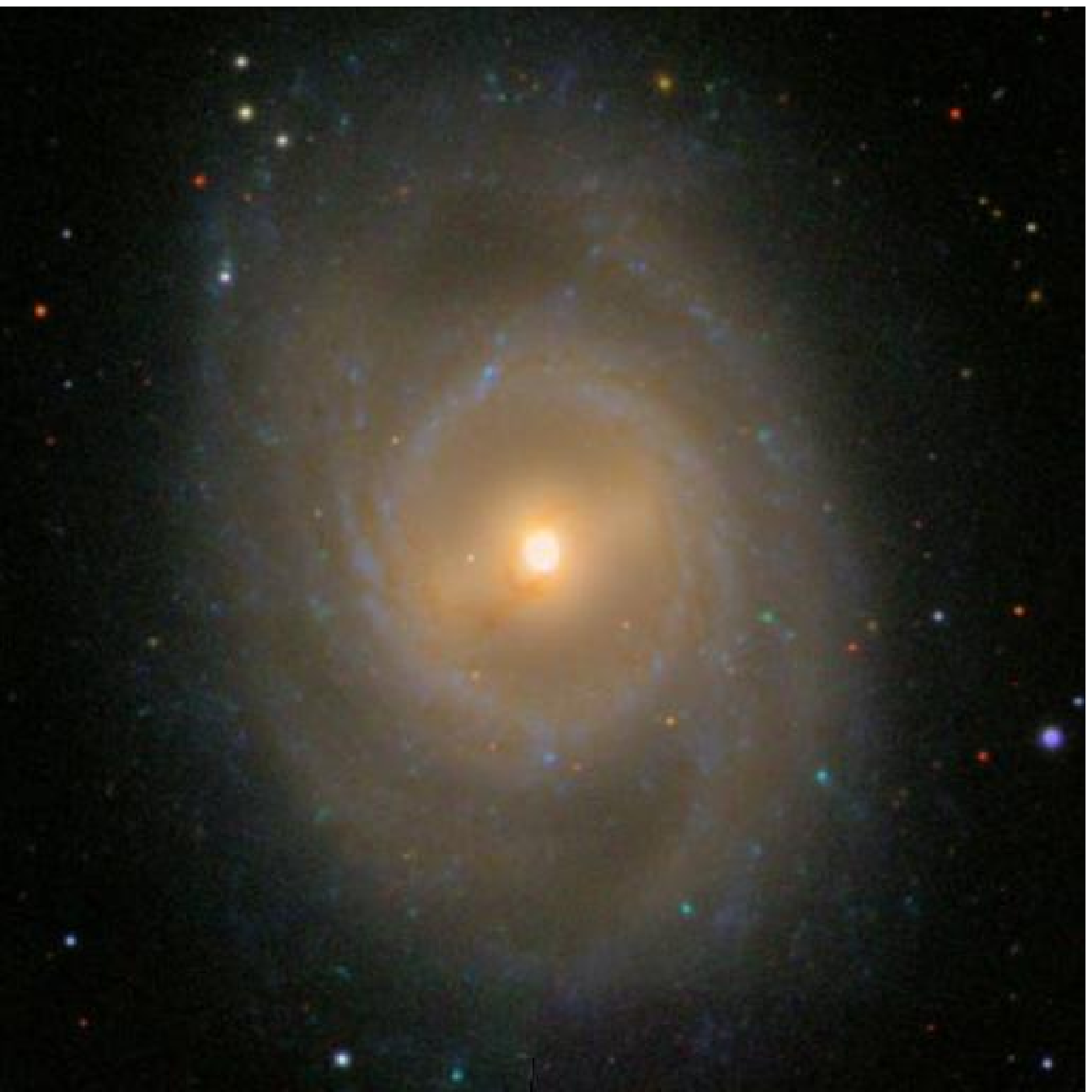}}
\caption{The galaxy NGC 3351, with a possible  bulge + long bar structure (SDSS image).}
\label{Fig:galaxies}
\end{figure}

Mart\'\i nez-Valpuesta \& Gerhard (2011, hereafter MG11) criticize the proposal
of the existence of a boxy bulge + long bar in the centre of the Milky Way 
(Hammersley et al.\ 2000; L\'opez-Corredoira et al.\ 2001, 2007 [hereafter L07]; Benjamin et al.\ 2005, 
Cabrera-Lavers et al.\ 2007 [hereafter C07]).
Finding possible problems in a hypothesis or alternatives to it is always an interesting exercise. 
Nonetheless, we find that MG11 is just a first step in an analysis that is still far from solving the problem
 of the structure in the inner 4 kpc of our Galaxy, and that leaves unexplained many observations related to 
 the possible existence of the long bar. The Basic gist of the article is as follows.
\begin{itemize}

\item The star-count maximum along a line of sight crossing a triaxial bulge/bar structure is not coincident with the major axis of the
structure.
That is indeed the case and has already been stated and discussed at length by our team (appendices A and B of L07; section 6 of C07).
Our conclusions, coincident with that of MG11, are that for a thick structure (the bulge), the difference 
between the angle of the real structure with respect to the apparent one in the plot of star-count maxima can be important, with
a systematic difference of up to $\approx 10^\circ $. The analyses by C07 of the angle of the
 thick bulge  take  this effect into account, and the structure with an apparent opening angle of 
 $\approx 25^\circ $ might indeed have a real inclination of $\approx 15 ^\circ $
(``opening angle'' refers to the orientation in the Galactic plane of structure with respect to the  Sun--Galactic centre line).  
In any case, this does not significantly affect  the hypothetical long thin bar (a triaxial bulge with 
axial ratios 1:0.25 in the plane  would give a maximum error of 100 pc in the
difference between maximum density and major axis; see appendix A of L07).

\item A bulge (developed from a bar after the second vertical buckling; Mart\'\i nez-Valpuesta et al.\ 2006)
with a twisted major axis of radius $\approx$4 kpc could reproduce the observed distribution 
of maxima in the plane of the red clump counts by C07 instead of the proposal by C07 and L07 of a shorter 
bulge + long bar with straight axes and a small 
angular difference between them. Indeed, from a purely morphological point of view,
 we are talking about the same thing under a different name. We could, for instance, say that the whole 
 structure in the centre of a galaxy like that shown in Fig.\ \ref{Fig:galaxies} is a bulge, a bar, or  
 a combined bar + bulge. What is evident is that this structure is thicker in the center and narrower
  at its extremes;  it is therefore well represented by a combination of thick bulge + long thin bar. Whether the name should be only a bulge or a bar or a bulge + bar is merely a question of semantics. The possible slight misalignment of the outer part of this structure with respect to its inner part ($\sim$20$^\circ $ in the Milky Way; see Ann 1995 for other galaxies) can also be interpreted as a single structure with twisted major axis.

\end{itemize}

There are further observational aspects not discussed by MG11 that need to be considered
 when analysing the possible existence of a long bar:

\begin{itemize}

\item Concerning morphology, one must also explain the measured thickness of the bar, not only 
the central position of the maxima of star counts. Figure 1a of MG11 shows a structure with a thickness at the tip of the bar at positive galactic longitude ($25^\circ <l<30^\circ $) of 4--5 kpc, whereas the 
thickness measured with the red clump method is  2--2.5 kpc (L07\footnote{We note that there is an erratum 
in L07, which says in the abstract and in the conclusions that the thickness of the bar is 1.2 kpc at 
$l=20^\circ $ and somewhat lower for higher longitudes; however, as correctly stated in the text of that article, this number is the $\sigma $ of a Gaussian distribution, equivalent to a dispersion of $\sigma \approx 0.5$ mag in the red clump distribution along the line of sight. Therefore, the axial ratios for the long bar are approximately 1:0.25:0.03.}). This difference in numbers is not too significant, but the important thing is that ``qualitatively'' we observe a much thicker bulge in the center ($l<15^\circ $) than in the outer parts ($l>15^\circ $),  and this observational result is not apparent in the proposal by MG11, which seems to maintain (judging from their fig.\ 1a) a similar axial ratio in the inner and the outer parts of their integrated bulge + bar structure.

\item With regard to asymmetries in the projected counts, one of the main motivations 
for positing the existence of the long bar was the fact that, within the plane ($b<2^\circ $), the star counts were far higher at positive galactic longitudes than negative longitudes with the same $|l|$, $b$, for $l< 30^\circ $ (L\'opez-Corredoira et al.\ 2001; section 2 of L07; fig.\ 20 of C07). This marked asymmetry vanishes at $b>2^\circ $ (L07). This indicates that the non-axysymmetrical structure must lie in the plane with a very low vertical thickness. MG11, based on Mart\'\i nez-Valpuesta et al. (2006), point out that their bulge becomes vertically thinner in the outer part, but they do not specify by how much. The bulge extends
between $b=-10^\circ $ and $+10^\circ $ in the inner parts and should be constrained 
within $|b|<2^\circ $ in the outer parts, to reproduce the star counts of C07 (their fig.\ 20). 
This is not shown by MG11. 

\item Regarding stellar populations, the division of a galaxy into several stellar components is not only
 a question of  visually identifying
substructures within the global morphology of the galaxy in question. It is also related to the separation of different populations with
different physical properties. This distinction is not rigid because, even in a given component like the thin disc or the bulge within $|l|<12^\circ $ in off-plane regions there are age and metallicity gradients, but an attempt is made to separate the major morphological groupings
according to their stellar populations.
In the case of the populations within galactocentric distances less than 4 kpc, there are important 
differences in the metallicity between the inner and outer parts (Gonz\'alez-Fern\'andez et al.\ 2008), 
so thinking about different populations associated with the bulge and the bar makes sense; alternatively, 
of course, one may posit  a unique component called the ``bulge'' with strong
outward metallicity gradient. In any case, the integrated bulge + bar structure proposed by MG11 cannot have a homogeneous stellar population
if it is to incorporate established observational evidence. 

\item There is gathering evidence for a star formation region (SFR) at the tips of the bar. 
It is well confirmed that there is a huge SFR in the
plane at $l\approx 27^\circ $ (L\'opez-Corredoira et al.\ 1999, Negueruela et al.\ 2011), the most 
prominent one in the Galaxy apart from the Galactic centre. This SFR marks the connection of the Scutum spiral
arm and the hypothetical long bar (L\'opez-Corredoira et al.\ 1999). It is composed of a burst of very young stars, nothing to do with the symmetric enhancements at the ends of the stellar bar, called {\it ansae}, or the ``handles'' of the bar/bulge (Mart\'\i nez-Valpuesta et al.\ 2007). 
This region is also detected in from methanol masers at 6.7 GHz (Green et 
al.\ 2011), together with the other huge SFR associated tentatively with the tip of the
 bar at negative longitudes ($l\approx -13^\circ $).
MG11 now claim now that the long thin bar is a thick boxy bulge extending to $R\approx 4$ kpc.
As argued by L\'opez-Corredoira et al.\ (1999), these kinds of star formation regions are observed in other galaxies that have a thin bar, but we do not know any case of a galaxy with the kind of boxy bulge  proposed by MG11.

\end{itemize}

From the point of view of methodology, we do not find the approach of MG11 to be the most appropriate. Provided that {\it all} the observational facts are reproduced, one should certainly use the simplest theoretical models rather than a complex model with many more free parameters (the principle of Occam's razor). MG11 claim that they cannot find a theoretical explanation for the existence of two misaligned triaxial structures, and that is their stated motivation for assert that something must be wrong with the interpretation of the observations of the Milky Way and other galaxies described here. This is a deductive standpoint (theorists telling observers what they should see). However, from an inductive standpoint (deriving theories from observations), which we find more appropriate, it is also possible to argue the need for changes in the theory rather than in the interpretation of observations. Two of us (Garz\'on \& L\'opez-Corredoira, in preparation) are currently working on theoretical models that allow the possibility of two misaligned bars/triaxial structures. 

Summing up, the model proposed by MG11 cannot replace the earlier proposal of a bulge + long bar, although
MG11's general consideration of an integrated boxy bulge and planar long bar might be possible
if a suitable model can be produced that explains all the relevant observational features.

%


\end{document}